\documentclass[5p,times,twoside,sort&compress]{elsarticle}
\pdfoutput=1
\usepackage{amsmath,amssymb}
\usepackage{color}
\usepackage{slashed}
\usepackage{tikz}
\usepackage{subfigure}

\newcommand{\skipthis}[1]{}
\newcommand{\imag}{\text{i}}
\renewcommand{\d}{\text{d}}
 \def\Eq#1{Eq.~(\ref{#1})}

\begin{document}
\title{Polyakov-Quark-Meson-Diquark Model for two-color QCD}
\author[label1]{Nils Strodthoff}
\author[label2,label3]{Lorenz von Smekal}

\address[label1]{Institut f\"{u}r Theoretische Physik, Universit\"{a}t Heidelberg, 69120 Heidelberg, Germany}
\address[label2]{Institut f\"{u}r Kernphysik, Technische Universit\"{a}t Darmstadt, 64289 Darmstadt, Germany}
\address[label3]{Institut f\"{u}r Theoretische Physik, Justus-Liebig-Universit\"{a}t Giessen, 35392 Giessen, Germany}

\begin{abstract}
We present an update on the phase diagram of two-color QCD from a chiral
effective model approach based on a quark-meson-diquark model using
the Functional Renormalization Group (FRG). We discuss the impact of
perturbative UV contributions, the inclusion of gauge field
dynamics via a phenomenological Polyakov loop potential, and the impact of
matter backcoupling on the gauge sector. The corresponding phase
diagram including these effects is found to be in qualitative
agreement with recent lattice investigations.   
\end{abstract}
\maketitle
\section{Introduction}
The understanding of the QCD phase diagram, in particular in regions
of intermediate chemical potentials, represents an enormous
theoretical challenge. The main obstacle to theoretical progress is
the sign-problem in QCD
\cite{Kogut:1999iv,Splittorff:2000mm,deForcrand:2010ys,Aarts:2013bla}. In
this situation it has become an important alternative to study finite
density effects in QCD-like theories with real fermion determinants
\cite{vonSmekal:2012vx}, as classified according to random matrix
theory by the Dyson index $\beta$ of their Dirac operators
\cite{Kogut:2000ek,Kanazawa:2011tt}. In the cases $\beta=1$, with
2-color QCD as a representative example, and $\beta=4$, as for QCD
with quarks in the adjoint representation or QCD with the gauge group
$G_2$ \cite{Holland:2003jy,Maas:2012wr}, the Dirac operator possesses
an additional antiunitary symmetry, which ensures the reality or even
positivity (for $\beta=4$ with Kramers degeneracy) of the fermion
determinant for a single quark flavor. In absence of such a symmetry,
for $\beta=2$ as in QCD, one is restricted to finite isospin density
\cite{Son:2000xc,Kogut:2004zg,Kamikado:2012bt}. Despite the fact that
such QCD-like theories differ in various important aspects from the
3-color world at finite baryon density, a better understanding of
their phase diagrams can provide insight into generic features of
finite density. At the same time they serve as benchmarks for quantum
field-theoretical continuum methods and model descriptions. In
particular, direct comparisons between functional continuum methods
and lattice simulations at finite density are possible in these
theories.

Two-color QCD has been studied within a number of different approaches
such as chiral perturbation theory \cite{Kogut:1999iv,Kogut:2000ek,Splittorff:2001fy}, 
random matrix theory \cite{Vanderheyden:2001gx,Klein:2004hv,Kanazawa:2011tt}, the NJL model \cite{Ratti:2004ra,Brauner:2009gu},
and on the Lattice \cite{Hands:2000ei,Kogut:2001na,Kogut:2001if,Hands:2006ve,Cotter:2012mb,Boz:2013rca}, 
see \cite{Strodthoff:2011tz} for a more extensive discussion of earlier approaches.
In this letter we expand on our previous Functional Renormalization
Group study of two-color QCD within the quark-meson-diquark (QMD) model
\cite{Strodthoff:2011tz}, where the model construction and the general
formalism was laid out, but where only the matter sector was taken
into account in the numerical results. Here we focus on the modeling
of gauge field dynamics in the form of a phenomenological Polyakov
loop potential \cite{Brauner:2009gu}. As compared to available
mean-field results we thereby also include the fluctuations due to collective
mesonic and baryonic excitations. At low baryon density, outside the
diquark condensation phase of two-color QCD, this extension is analogous
to that of the Polyakov-quark-meson model for the QCD phase diagram
\cite{Schaefer:2007pw} when mesonic fluctuations are included
 \cite{Herbst:2010rf, Skokov:2010uh,
    Skokov:2010wb,Herbst:2012ht,Herbst:2013ail,Haas:2013qwp}. 
With diquark condensation and diquark fluctuations, however, this will
include the region of high baryon density in the phase diagram of
two-color QCD and thus allow a more detailed comparison with recent 
lattice results \cite{Boz:2013rca,Cotter:2012mb}. 

\section{Theoretical Background}
In this section we review the essentials of two-color QCD and
its effective Polyakov-quark-meson-diquark (PQMD) model
description. We furthermore introduce the necessary basics of the
Functional Renormalization Group approach and the corresponding flow
equations for the effective potential of the model in the leading
order derivative expansion. 

\subsection{PQMD model for two-color QCD}
\label{subsec:pqmd}

\begin{figure}[ht]
\centering
\begin{tikzpicture}[->,node distance=1.5cm,auto,thick,main node/.style={draw,font=\bfseries},scale=.1]
  \node (A) {$\textit{SU}(2N_f)$};
  \node (B) [below of=A, left of=A, node distance=1.8cm] {$\textit{Sp}(N_f)$};
  \node (C) [below of=A, right of=A,node distance=1.8cm] {$\textit{SU}(N_f)_L\!\times\! \textit{SU}(N_f)_R\!\times\! U(1)_B$};
  \node (D) [below of=A, node distance=3.6cm] {$\textit{SU}(N_f)_V\times U(1)_B$};
  \node (E) [below of=D, node distance=1.8cm] {$\textit{Sp}(N_f/2)_V$};
  \draw (A) to node {$\mu>0$} (C);
  \draw (A) to node[swap] {$m_q>0$} (B);
  \draw (B) to node {} (D);
  \draw (C) to node {} (D);
  \draw (D) [dashed,text width=1.8cm] to node {diquark cond.} (E);
  \path (D) to node [swap] {$\frac{N_f(N_f-1)}{2}$ GBs} (E);
\end{tikzpicture}
\caption{Patterns of symmetry breaking in two-color QCD with $N_f$ flavors of fundamental quarks ($\beta=1$)}
\label{fig:symbr1}
\end{figure}
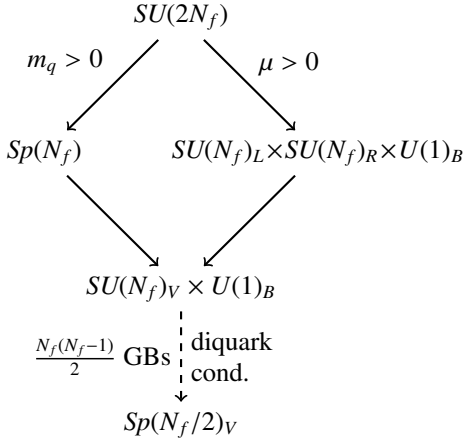

The key to understanding the special properties of two-color QCD
is its enlarged flavor symmetry, which is in turn based on the
pseudo-reality of the $\textit{SU}(2)$ fundamental representation. In a theory
with $N_f$ degenerate quark flavors the enlarged flavor symmetry group is given by $\textit{SU}(2N_f)$
which contains the usual flavor and baryon
number $\textit{SU}(N_f)_L\times \textit{SU}(N_f)_R\times U(1)_B$ symmetries as subgroup. 
Obviously, the enlarged flavor symmetry also changes the
pattern of chiral symmetry breaking; an explicitly or spontaneously
generated Dirac mass term breaks the enlarged $\textit{SU}(2N_f)$ to the
symplectic group $\textit{Sp}(N_f)$, whereas the inclusion of a chemical
potential breaks it to $\textit{SU}(N_f)_L \times \textit{SU}(N_f)_R \times U(1)_B$. In presence of both, the residual symmetry is given by
the common $\textit{SU}(N_f)_V \times U(1)_B$ subgroup of the two.
In the diquark condensation phase this symmetry gets broken spontaneously to
$\textit{Sp}(N_f/2)$ and correspondingly $N_f(N_f-1)/2$ Goldstone bosons
occur. For asymptotically large chemical potentials chiral symmetry
gets (partially) restored to $\textit{Sp}(N_f/2)_L\times \textit{Sp}(N_f/2)_R$. It is a
special property of the 2-flavor theory that this leads to a complete
restoration of the chiral $\textit{SU}(2)_L\times
\textit{SU}(2)_R$ symmetry at asymptotically large
chemical potentials. The symmetry breaking patterns in two-color QCD
with $N_f$ degenerate flavors of fundamental quarks are summarized in
Fig.~\ref{fig:symbr1}.  

In the following we will concentrate on the case of two flavors where
the breaking $\textit{SU}(4)\to \textit{Sp}(2)$ is locally the same as
the simple vector-like breaking of $\textit{SO}(6)\to
\textit{SO}(5)$. The corresponding five (pseudo-)Goldstone bosons are
identified with the three pions plus a scalar bosonic diquark/antidiquark
pair, which is thus  degenerate with the pions at vanishing chemical
potential. These diquarks play a dual role as pseudo-Goldstone
bosons and as the lightest baryonic degrees of freedom in the
theory. In this case, diquark condensation simply corresponds to
$\textit{SU(2)}_V\times \textit{U(1)}_B\to \textit{Sp(1)}_V\simeq
\textit{SU(2)}_V$, i.e.\ to the spontaneous breaking of the
$\textit{U(1)}_B$ for baryon number conservation.  
Most importantly, the pattern of symmetry
breaking is correctly reproduced in a quark-meson-diquark model, as an
effective model of quarks, mesons and diquarks. 

 In the case of two quark flavors it is described by the (Euclidean) Lagrangian \cite{Strodthoff:2011tz}
\begin{equation}
\label{eq:pqmdlagrangian}
\begin{split} \mathcal{L}_\text{PQMD}=&\bar{\psi}\left(\slashed{D}+h(\sigma+\imag\gamma^5\vec{\pi}\vec{\tau})-\mu
    \gamma^0\right)\psi\\
  &+\frac{h}{2}\left(\Delta^* (\psi^T C \imag\gamma^5\tau_2T_2 \psi)+ \Delta(\psi^\dagger C \imag\gamma^5\tau_2T_2\psi^*)\right)\\
  &+\frac{1}{2} (\partial_\mu \sigma)^2+\frac{1}{2} (\partial_\mu \vec
  \pi)^2  +
  V(\vec \phi)\\
  &+\frac{1}{2} (\partial_\mu-2\mu \delta_{\mu
    0})\Delta(\partial^\mu+2\mu \delta^\mu_0)\Delta^*\ +\mathcal{U}_\text{Pol},
\end{split}
\end{equation}
with Yukawa coupling $h$, and $\tau_i$ denoting Pauli matrices in
flavor space; $T_i=\frac{\sigma_i}{2}$ are the $\textit{SU}(2)$  color
generators and $C=\gamma^2 \gamma^0$ is the charge conjugation matrix
in spinor space. We furthermore define the vector $\vec\phi
=(\sigma,\vec \pi,\text{Re}\,\Delta,\text{Im}\,\Delta)$ of meson and
diquark fields which transforms as a vector under the enlarged
$O(6)\simeq \textit{SU}(4)$ flavor symmetry. The color covariant
derivative is given by $D_\mu= \partial_\mu + \imag A_\mu$ with a
constant background gauge field
$A_\mu=\delta_{\mu 0}A_0$ in the Polyakov gauge, i.e. for
$\textit{SU}(2)$ simply with $A_0=T^3 a_0$. The Polyakov loop whose
thermal expectation value serves as an order parameter for confinement
in the pure gauge theory is thus represented as 
\begin{equation}
\label{eq:pldefinition}
\Phi\equiv\tfrac{1}{2}\text{tr}_c e^{\imag \beta
  A_0}=\cos\left(\tfrac{\beta a_0}{2}\right). 
\end{equation}
While one could employ Polyakov loop potentials from lattice
simulations or functional continuum methods \cite{Haas:2013qwp} in the
future, here we present results for a phenomenological Polyakov loop potential
\cite{Brauner:2009gu} of the form 
\begin{equation}
\label{eq:upol}
\mathcal{U}_\text{Pol}(\Phi;T,T_0)=-b T [24 \Phi^2 e^{-a/T}+\log(1-\Phi^2)],
\end{equation}
which is a 2-color variant of the commonly used 3-color logarithmic
Polyakov loop potential \cite{Fukushima:2008wg,Fukushima:2003fw}. The
deconfinement transition itself is fixed by the parameter $a$ which is
related to the critical temperature $T_0$ of the pure gauge theory as
$a=T_0 \log 24$, whereas a strong coupling expansion relates $b$ to
the string tension $\sqrt{\sigma}$ via $b=(\sigma/a)^3$. The parameter
$b$ determines the mixing between chiral and deconfinement
transition and can be used to adjust the pseudocritical
temperature for the chiral transition relative to 
the deconfinement transition. It is typically chosen such that the
two crossovers coincide \cite{Fukushima:2008wg}. Here we simply fix
$b=(\sigma/a)^3$ and adjust $T_0$ with $N_f$ and $\mu$ as described
below. The rational for this adjustment is to account for the implicit
feedback of the matter sector on the gluodynamics and hence the
Polyakov loop potential \cite{Schaefer:2007pw}. This includes sea
quark effects on the gluonic correlations, for example, in
contradistinction to the valence quark contributions as here described
explicitly by the fermionic flow.    

The critical temperatures of pure $\textit{SU}(N_c)$ gauge theories have been
well-investigated on the lattice, see \cite{Lucini:2012wq} and the
references therein. In units of the string tension they are very well
described by the corresponding value in the large $N_c$ limit plus
a $1/N_c^{2}$ correction term, all the way down to $N_c=2$
\cite{Lucini:2003zr}.  For $\textit{SU}(2)$ this yields
$T_{c}/\sqrt{\sigma}=0.7092(36)$ \cite{Lucini:2012wq} which 
corresponds to $T_c= 312$~MeV in physical units assuming a string
tension with $\sqrt{\sigma}=440$~MeV. This fixes $T_0(N_f=0,
\mu=0) = T_c $.

We may generally relate couplings $\alpha_T$ and $\alpha_0$ at sufficiently
close-by  temperature scales $T$ and $T_0$  assuming a
logarithmic dependence \cite{deForcrand:2001nd} of the form 
\begin{equation}
\ln (T/T_0)  = a \, \bigg(\frac{\alpha_0}{\alpha_T} -1
\bigg)  \label{logTscale}  
\end{equation}
with some nonperturbative coefficient $a$ which depends on $N_f$ and
$N_c$ (for $N_f=0$ of the order $1/N_c$). In order to include Debye
screening effects we consider the {\em effective charge}
$\alpha_\mathrm{eff}(p)$ in the plasma \cite{Lebellac:1996} at 
the soft scale $p\sim gT$ as in  \cite{Schaefer:2007pw},
\begin{equation}
\alpha_\mathrm{eff}(gT) = \frac{\alpha_T}{1+b(\mu/T)}\, ,
\end{equation}
where $b(\mu/T) \equiv m_D^2/(gT)^2$ is given by the Debye mass
$m_D$ per $p\sim gT$.  At one-loop level it would be \cite{Lebellac:1996}
\begin{equation}
b(\mu/T) = \frac{N_c}{3} +\frac{N_f}{12}
\bigg( 1 + \frac{3}{\pi^2} \frac{\mu^2}{T^2} \bigg) \, .
\end{equation}
The Debye mass increases with $\mu$ and hence the effective charge
decreases. The simplest way to include Debye screening thus is to
consider lines $T(\mu)$ at constant $\alpha_\mathrm{eff}$, with
$T_0 \equiv T(0)$, $b_0 \equiv b(0)$  and (\ref{logTscale}) these are
obtained as, 
\begin{equation}
\ln\big(T(\mu)/T_0\big) =  \frac{b_0-b(\mu/T)}{1+b(\mu/T)} \, a \, ,
\end{equation}
where $a$ is the $N_c$ and $N_f$ dependent but $\mu$-independent
nonperturbative coefficient from Eq.~(\ref{logTscale}). If we expand
\begin{equation}
\label{eq:bmuexpansion}
b(\mu/T) = b_0 + b_1 \mu^2/T^2 + \dots
\end{equation} 
the leading logarithmic behavior of
$T(\mu)$ near $\mu=0$ becomes
\begin{equation}
\ln\big(T(\mu)/T_0\big) =  - \frac{a b_1}{1+b_0} \frac{\mu^2}{T_0^2}
\, .
\label{Tofmu}
\end{equation}
We can test this simple argument with the critical temperatures of the
pure $\textit{SU}(N_c)$ gauge theories: Using $N_f=0$, $\alpha \sim
1/N_c $ and $ b \sim N_c$ in the large $N_c$ limit, one concludes that
the effective charge decreases as $1/N_c^2$. Because the number of
gluons grows with $N_c^2$, the assumption that $N_c^2
\alpha_\mathrm{eff} = $ const.~in this case, together with $a\sim
1/N_c$ in (\ref{logTscale}), yields   
\begin{equation}
\ln\big(T_c(N_c)/T_c^\infty\big) =  \, \frac{c}{N_c^2},  \label{expfit}
\end{equation}
with some constant $c$.
Fitting the lattice data for $N_c= 2,\dots 8$ as collected in
\cite{Lucini:2012wq} to this two parameter form works quite well. For
comparison, we obtain with this form
$T_c^\infty /\sqrt{\sigma} = 0.5962(16)$ with a reasonable 
$\chi^2/\mathrm{d.o.f.} = 1.27$, as compared to  
\begin{equation} 
 T_c/\sqrt{\sigma} = 0.5949(17) + 0.458(18)/N_c^2 \,
 , \label{Lucinifit} 
\end{equation}
from \cite{Lucini:2012wq}
with $\chi^2/\mathrm{d.o.f.} = 1.18$.
If we exclude the $N_c=2$ value for not being
close enough to the large $N_c$ limit, we obtain $T_c^\infty
/\sqrt{\sigma} = 0.5952(24)$ and our fit with (\ref{expfit}) is
practically indistinguishable from (\ref{Lucinifit}) of
Ref.~\cite{Lucini:2012wq} for $N_c\ge 3$. It is thus consistent with
general large-$N_c$ arguments at this order \cite{Toublan:2005rq}.

An analogous argument also applies when varying the number of flavors
$N_f$. In order to model the density dependence of Debye screening we
therefore simply replace the parameter $T_0$ in the Polyakov loop
potential (\ref{eq:upol}) by the line $T_0(N_f,\mu) $ of {\em constant
effective charge} with $T_0(0,0) = T_c =312$ MeV (for $N_c=2$
here). From (\ref{Tofmu}) at the leading order in $\mu^2/T_c^2$ this line will
hence be of the form, with new constants $a$ and $b$,    
\begin{equation}
\label{eq:T0scaling}
T_0(N_f,\mu)=T_{c} \,\exp\left(-a N_f \bigg(1+ b
\frac{\mu^2}{T_{c}^2}\bigg)\right)\, .
\end{equation}
The non-perturbative coefficient $a$ herein should first be fixed such
that the deconfinement temperature at vanishing chemical potential matches 
(suitably extrapolated) lattice results. Since 
$\textit{SU}(2)$ simulations with comparably light dynamical quarks
are phenomenologically less relevant than those of real QCD
thermodynamics, they have received less attention and results are
therefore rather limited. As an orientation we
use the value for the deconfinement crossover temperature of
around 217~MeV obtained from simulations with two degenerate flavors
of dynamical Wilson quarks \cite{Boz:2013rca,Cotter:2012mb},
albeit with masses considerably above their physical counterparts in
QCD. Because the transition temperature is expected to further
decrease with decreasing quark masses, we employed a value of $a=0.19$
corresponding to $T_0(2,0)= 212$~MeV which will then lead to a
deconfinement crossover temperature of around 200~MeV. 

A reasonable way to fix the second
non-perturbative coefficient $b$ in Eq.~(\ref{eq:T0scaling}) would be
to match the curvature of the pseudocritical line extracted from the
lattice a posteriori. Again due to a lack of suitably accurate lattice
data for two-color QCD we have only investigated the impact of different
parameter values for $b$ in a more exploratory fashion for now. In the 
following section we will simply compare results with
$b=0$, $b=2.6$ and $b=5.2$ to exemplify the impact of Debye screening in the 
unquenching effects from the matter sector on the deconfinement transition at
finite density. 

\subsection{Functional Renormalization Group}
The Functional Renormalization Group is a powerful non-perturbative
tool for calculations in quantum field theory and statistical
physics. Here we employ the approach pioneered by Wetterich
\cite{Wetterich:1992yh} with a so-called effective average action as
the central object, see
\cite{Berges:2000ew,Polonyi:2001se,Pawlowski:2005xe,Schaefer:2006sr,Gies:2006wv,Braun:2011pp}
for general introductions. The FRG aims at computing the
full quantum effective action by relating a classical or microscopic
bare action at the ultraviolet cutoff scale $\Lambda$ to the
corresponding  
average action at some lower scale $k$, the scale-dependent
analogue of the effective action. This RG scale $k$
introduced by an infrared regulator is then successively lowered 
which yields the evolution of the scale-dependent
effective average action with the RG scale $k$ or, correspondingly,
with  $t=\log k/\Lambda$ as described by the exact flow equation 
\begin{equation}
\label{eq:floweq}
\partial_t \Gamma_k=\frac{1}{2}\text{STr}\left\{[\Gamma_k^{(2)}+R_k]^{-1}\partial_t R_k\right\},
\end{equation} 
which assumes the form of a 1-loop equation, however, involving full
(scale- and field-dependent) propagators. Here $\Gamma^{(2)}$ denotes
the second functional derivative of the effective average action with
respect to the fields and the supertrace involves a trace both over
momentum space and internal indices and includes an additional minus
sign in the fermionic subsector. As the flow equation
(\ref{eq:floweq}) can rarely be solved exactly truncations are
required. Here we employ the leading order derivative expansion in
which only a scale-dependent effective potential is taken into
account. Thus the Ansatz for the effective average action simply reads,
in terms of the Lagrangian (\ref{eq:pqmdlagrangian}), 
\begin{equation}
\Gamma_k=\int \d^4 x \mathcal{L}_{PQMD}|_{V(\phi)\to U_k(\rho^2,d^2)-c\sigma},
\end{equation}
where $\rho^2=\sigma^2+\vec \pi^2$ and $d^2=\Delta^* \Delta$ denote
the two $\textit{SU}(2)\times U(1)_B$ invariants and the $c\sigma$
term represents an explicit breaking, which is taken into account at
the end of the flow. It is crucial to consider an Ansatz for the
scale-dependent effective potential $U_k$ which is a genuine function
of the two independent invariants $\rho^2$ and $d^2$ as the potential
at finite chemical potential is only required to be consistent with
the reduced symmetry $\textit{SU}(2)\times U(1)_B$ instead of the full
enlarged flavor $\textit{SU}(4)$ at vanishing chemical potential. This
Ansatz is consistent with the full $\textit{SU}(4)$ symmetry at
$\mu=0$ as one can recast it as a function of a single variable
$\phi^2=\rho^2+d^2$ again. Employing 3-dimensional
  analogues of the LPA-optimized regulator functions
  \cite{Litim:2001up} which are commonly used in finite-temperature
  applications \cite{Litim:2006ag},  
$R_{k,B}=(k^2-\vec p^2)\Theta(k^2-\vec p^2)$ and $R_{k,F}=\imag \slashed {\vec {p}}(-1+k/|\vec p|)\Theta(k^2-\vec p^2)$,
for bosonic and fermionic fields respectively, the flow equation for the effective potential takes the form \cite{Strodthoff:2011tz}
\begin{multline}
\label{eq:fullflowfinalpqmd}
  \partial_t U_k=
  \frac{k^5}{12\pi^2}\left\{\frac{3}{E^\pi_k}\left(1+2 n_b(E^\pi_k;T)\right)\right.\\
    \left.+\sum_{i=1}^3  \frac{3 z_i^4-\alpha_1
   z_i^2+\alpha_0} {(z_{i+1}^2-z_i^2) (z_{i+2}^2-z_i^2)}
    \frac{1}{z_i} \left(1+2 n_b(z_i;T)\right)\right.\\
   \left.-\sum_{\pm}\frac{8}{E^\pm_k}\left(1 \pm
      \frac{\mu} {\epsilon_k}\right) 
   \, \left( 1 - 2 n_q(E_k^\pm;T,\Phi) \right) 
\right\}\ , 
\end{multline}
where $E^\pi_k=\sqrt{k^2+2 \partial U_k/\partial \rho^2}$,
$E^\pm_k=\sqrt{h^2 d^2+(\epsilon_k\pm\mu)^2}$ and
$\epsilon_k=\sqrt{k^2+h^2\rho^2}$. The quantities $z_i$ in the
sigma-diquark sector denote the roots of a cubic polynomial in $p_0^2
= - z^2$ with coefficients $\beta_i$. These together with
the coefficients $\alpha_i$ of the corresponding quadratic polynomial in the
numerator are listed explicitly in
\cite{Strodthoff:2011tz,vonSmekal:2012vx}.
They all depend on the renormalization scale, the field invariants
$\rho^2$ and $d^2$, on the chemical potential and the
derivatives of the scale dependent effective potential. The Polyakov
loop enhanced fermion occupation numbers are given by  
\begin{equation}
n_{q}(E;T,\Phi)=\frac{1+\Phi e^{E/T}}{1+2\Phi e^{E/T} +e^{2E/T}}\, .
\end{equation}
and reduce to the usual Fermi-Dirac distribution for $\Phi=1$,
whereas $n_b(E;T)=1/(\exp(E/T)-1)$ denotes the Bose-Einstein
distribution function.

In our approach the gauge field $a_0$ is treated as a background field. The integration of \Eq{eq:fullflowfinalpqmd} yields an effective potential as function of $\rho^2$, $d^2$ and $a_0$ which is then minimized with respect to all three variables to obtain chiral and diquark condensates and the expectation value of the Polyakov variable $\Phi$ as a function of temperature and chemical potential.

In numerical calculations it is of course important to remember the
range of validity of the approach. For a given UV cutoff
$\Lambda $ the assumption of a temperature- and chemical-potential-independent bare action in the UV, for example,
severely restricts the accessible range of temperatures and/or chemical
potentials. This is most easily seen in the case of finite temperature
where the flow starts to deviate from the vacuum flow only at around
$k\approx 2\pi T$. This restricts the allowed temperature range at
a fixed UV cutoff to values below $T \sim \Lambda/(2\pi)$. The only way
of enlarging this range is to augment the model result
with the expected perturbative behavior, which then also ensures
thermodynamic consistency. In fact one may understand these
additional perturbative UV contributions as being necessary to describe
the thermodynamics of the microscopic model at the UV cutoff
scale. To achieve this one can integrate the purely thermal flow,
i.e. the difference between finite temperature and vacuum flow, from the UV
cutoff scale $\Lambda $ to infinity. The result
is then added to the UV potential before integrating the flow equation
(\ref{eq:fullflowfinalpqmd}). Obviously this gives rise to temperature-
and chemical-potential-dependent initial conditions in the UV. Note
that such a procedure can in general not merely modify the thermodynamics but
affect the phase structure itself. Here we implement this improvement
only in the fermionic fluctuations for which the purely thermal flow reads, 
\begin{equation}
\label{eq:perturbativeuvcorr}
\partial_t U_k^{(T,\mu)}-\partial_t U_k^{(T=0,\mu)}= \frac{k^5}{3\pi^2}\sum_\pm\frac{4}{E^\pm_k}\bigg(1 \pm
      \frac{\mu} {\epsilon_k}\bigg) \,  n_q(E_k^\pm;T,\Phi) \, ,
\end{equation}
see also \cite{Litim:2006ag} for a discussion of
  purely thermal flows.
As compared to previous studies
\cite{Braun:2003ii,Skokov:2010uh,Skokov:2010wb}, which included
analogous but field-independent UV contributions to ensure a proper 
Stefan-Boltzmann limit, we include the full field dependence here. A
particular simplification in the fermionic sector thereby is that
the right hand side of the corresponding flow is independent of the
effective potential and can be straightforwardly integrated.

\subsection{Numerical procedure}
For a fixed value $a_0$ of the background gauge field the flow
equation (\ref{eq:fullflowfinalpqmd}) for the effective potential was
solved on a two-dimensional grid in field space as in
Ref.~\cite{Strodthoff:2011tz} thereby retaining the full field dependence
of the equation. As the gauge field is treated simply as a background
field in our approach, the full three-dimensional effective potential
as a function of the invariants $\rho^2,d^2$ and $a_0$ in the IR is
obtained by combining results from runs with different values of
$a_0$. In this way one obtains a discretized IR potential which can be
interpolated for example using cubic splines and which is subsequently
minimized. Not only does this provides a very efficient way of
minimizing the full two-dimensional effective potential in the infrared,
using relatively few of the expensive evaluations of the flow
equation, but it also allows to conveniently extract its
derivatives at the minimum which can then be used to define crossover
criteria as discussed below.  

\section{Results}
\subsection{Impact of thermal UV contributions}

\begin{figure}[t]
    \includegraphics[width=0.97\columnwidth]{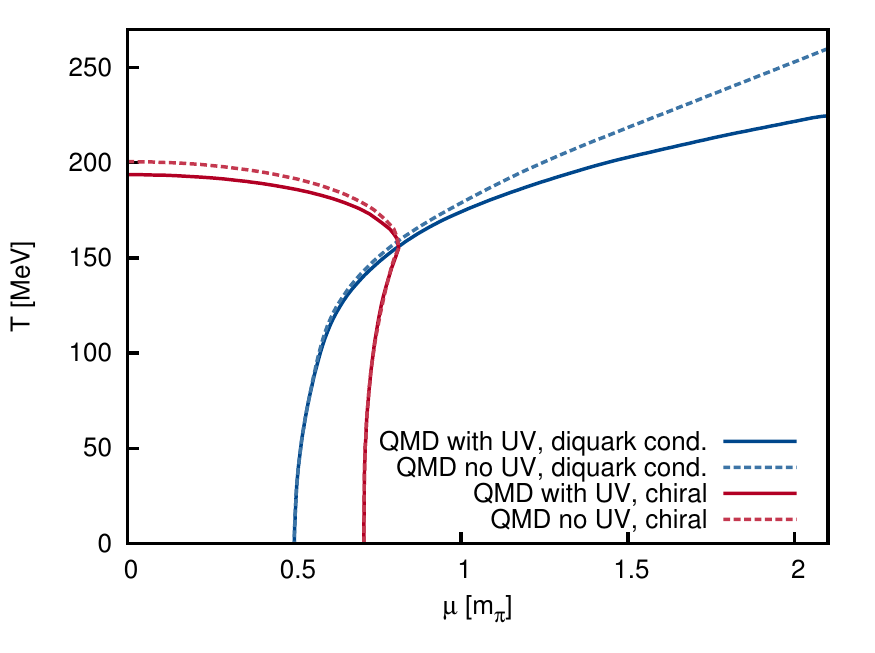}
    \caption{Comparison of FRG results for the QMD model phase diagram
      with and  without the thermal UV contributions
      (\ref{eq:perturbativeuvcorr}) to the fermionic flow. Chiral crossover lines (half-value of the chiral
  condensate) are depicted in red, the second order phase boundary of
  the diquark condensation phase found at small temperatures and large
  chemical potentials in blue.}   
    \label{fig:qmduvcomparison}
   \end{figure}

The effect of the perturbative UV contributions discussed in the
previous section is seen in Fig.~\ref{fig:qmduvcomparison}
where we compare the full FRG result for the QMD model phase diagram
with these thermal UV contributions from
(\ref{eq:perturbativeuvcorr}) to the corresponding result of
Ref.~\cite{Strodthoff:2011tz} without them. The phase
  diagram shows a phase of broken chiral symmetry at small
  temperatures and chemical potentials whereas for large chemical
  potentials one finds a diquark condensation phase signaled  
by a nonvanishing diquark condensate, see \cite{Strodthoff:2011tz} for
a more detailed discussion of the QMD model phase diagram. The
inclusion of UV contributions leads to a slight 
suppression of the chiral condensate at larger temperatures which 
consequently shifts the chiral crossover line to somewhat 
lower temperatures. A similar effect is observed for values of the
chemical potential $\mu$ above the onset of diquark condensation at half the
pion mass $m_\pi$. The boundary of the diquark condensation phase gets
pushed towards lower temperatures more and more as $\mu$ is further
increased. It is important to note, however, that in contrast to  
corresponding mean-field calculations \cite{Strodthoff:2011tz}, for which
the inclusion of the full thermal contributions lead to the appearance
of a tricritical point along the diquark condensation phase boundary,
with fluctuations this boundary remains to be of second order
throughout the entire parameter range investigated here.

\subsection{Vanishing chemical potential and crossover criteria\label{sec:criteria}}

\begin{figure}[t]
    \includegraphics[width=0.97\columnwidth]{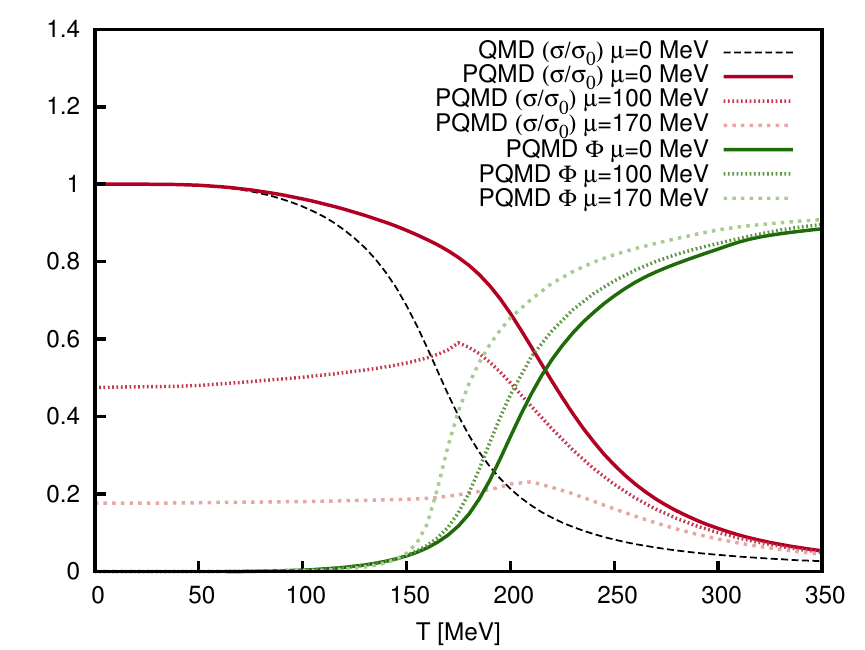}
    \caption{Chiral condensate (PQMD and QMD model) and Polyakov loop as function of temperature for vanishing and non-vanishing chemical potential.}
    \label{fig:condensatesmu0}
\end{figure}

We start our discussion of the PQMD model results with the case of
vanishing chemical potential where the effective potential is still
required to show the full enlarged $\textit{SU}(4)\simeq
\textit{SO}(6)$ symmetry and hence the calculation with the $O(6)$
symmetric effective potential coincides with the full solution. In
\cite{Strodthoff:2011tz} it was verified by an analysis of the
critical exponents that the finite temperature transition was
consistent with the expected $O(6)$ universality class corresponding
to a symmetry breaking pattern $\textit{SU}(4)\to \textit{Sp}(2)$ or
isomorphically $\textit{SO}(6)\to \textit{SO}(5)$. This will still
hold in the present case since the critical physics is 
governed by the bosonic matter sector, at least as long
as the Polyakov loop is taken as a background  (mean-)field without
dynamical matter feedback on gauge field fluctuations.    

Fig.~\ref{fig:condensatesmu0} shows the temperature dependence of the
chiral condensate and the Polyakov loop as quasi-order parameters for the
chiral and deconfinement transitions. For comparison we also
include the chiral condensate obtained from a pure QMD model
calculation corresponding to a fixed value of $a_0=0$. At this point
the main effect of the Polyakov loop on the chiral condensate is to
shift the chiral transition to larger temperatures. As both the chiral
and the deconfinement transition turn into crossovers for finite quark
masses, the corresponding transition temperatures are not uniquely
defined. We therefore compare up to three different definitions and
use the resulting variations as indications for the widths of the two
crossovers. The simplest one is the point where the order parameter
reaches half of its value at $T=\mu=0$. While this does not even
define a proper pseudo-critical temperature, it turns out to be a
useful measure for chiral restoration in the diquark condensation
region, as discussed in Subsection \ref{subsec:fullflow}, 
where there is no pseudo-critical line.
A second commonly used definition is the inflection point of the order
parameter along the temperature axis, i.e. the extremum of its 
temperature derivative which is readily computable, in principle.
This can become increasingly difficult, however, in regions where the slope
of the order parameter is nearly constant, and it fails entirely of
course when there is no inflection point as for the chiral condensate in the
diquark condensation region for $\mu > m_\pi/2$,
c.f.~Fig.~\ref{fig:condensatesmu0}.  
Finally as a third criterion we use the maxima of the
corresponding susceptibilities, i.e. the chiral and the Polyakov loop
susceptibility, which are easily accessible by taking the appropriate
second derivatives of the effective potential with respect to the order
parameters. The maxima of the susceptibilities define proper
pseudo-critical lines \cite{Pelissetto:2000ek} and are thus the
probably most natural choices from the point of view of critical
phenomena.  

\begin{figure}[t]
    \includegraphics[width=0.97\columnwidth]{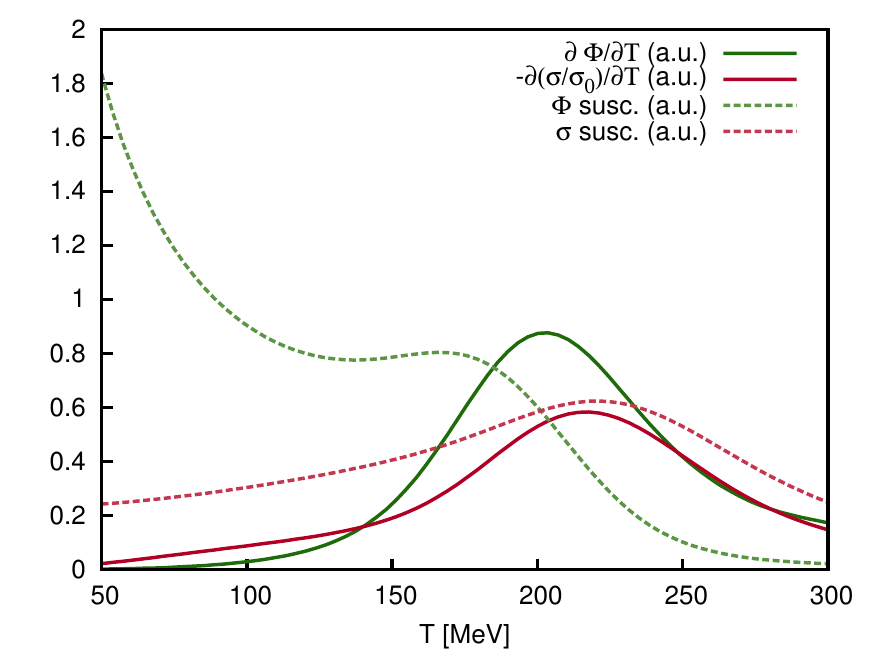}
    \caption{PQMD model at $\mu=0$: Comparison of crossover criteria.}
    \label{fig:derivativesmu0}
\end{figure}

The corresponding temperature derivatives and
susceptibilities are shown in Fig.~\ref{fig:derivativesmu0} and the
associated crossover temperatures are compiled in
Table~\ref{tab:Tcs}. These values should not be taken too literally as
quantitative predictions because they show some 
sensitivity to the parameters, especially to the adjusted sigma
mass. As mentioned above, the deconfinement crossover temperature
from the inflection point is somewhat below the central value but
still within errors of the corresponding lattice result of 217(23)~MeV
\cite{Cotter:2012mb}, which was obtained from simulations with
considerably larger quark masses, however.

\subsection{$O(6)$ symmetric effective potential}

In this subsection we address the calculation with an $O(6)$ symmetric
effective potential depending on the single invariant
$\phi^2=\rho^2+d^2$. For $\mu=0$ the solution coincides with the full
solution, but it will deviate from that at finite $\mu$. Nevertheless, it
represents a good approximation to the full solution at least for small
chemical potentials where the $O(6)$ symmetry still holds
approximately. As discussed in \cite{Strodthoff:2011tz} this
calculation closely resembles the corresponding Polyakov-quark-meson
model calculations for 3-color QCD \cite{Herbst:2010rf, Skokov:2010uh,
  Skokov:2010wb,Herbst:2012ht,Herbst:2013ail}. On one hand, the
only difference in the chiral sector is the a larger number of
would-be Goldstone bosons, five here instead of the three pions for the case of two
light flavors in QCD, see \cite{Kogut:2000ek} for a discussion
of the Goldstone spectrum. Two of these five pseudo-Goldstone bosons couple
to the chemical potential in a way analogous to 
the coupling of charged pions to an isospin chemical potential
QCD, see \cite{Kamikado:2012bt} for a detailed discussion of the
relation between two-color QCD at finite baryon density and QCD at
finite isospin density. On the other hand the gauge sectors in the two
theories are of course fundamentally different. Despite these
differences the corresponding phase diagrams turn out to share the same  
qualitative behavior.

\begin{figure}[!t]
    \includegraphics[width=0.97\columnwidth]{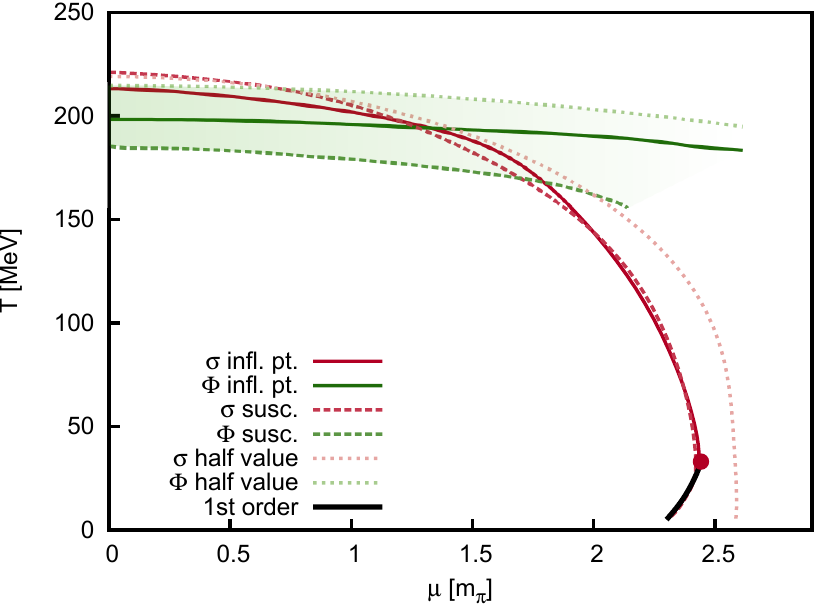}
    \caption{PQMD model phase diagram for an $O(6)$ symmetric effective potential and constant $T_0$. Chiral crossover lines are depicted in red, deconfinement crossover lines in green for three different crossover criteria, cf.\ the discussion in Sec.~\ref{sec:criteria}, and first order transitions in solid black.}
    \label{fig:pqmdd0nomu}
   \end{figure}

\begin{table}[!ht]
\begin{center}
 \begin{tabular}{|c|c|c|}
\hline  
  criterion& $T_c^\text{chiral}$ [MeV]& $T_c^\text{deconf.}$ [MeV]\\
  \hline
  \hline
  half-value&219.1&214.4\\
  inflection pt. &213.0&198.2\\
  susceptibility &221.1&185.5\\
  \hline
 \end{tabular}
\end{center}
\vspace{-.4cm}
\caption{Chiral and deconfinement crossover temperatures at $\mu=0$ for pion decay constant $f_\pi=76$~MeV, a physical pion mass $m_\pi=138$~MeV defined via the onset of the onset at vanishing temperature and a sigma (screening-)mass of 551~MeV.} 
\label{tab:Tcs}
\end{table}

\begin{figure}[!t]
    \includegraphics[width=0.97\columnwidth]{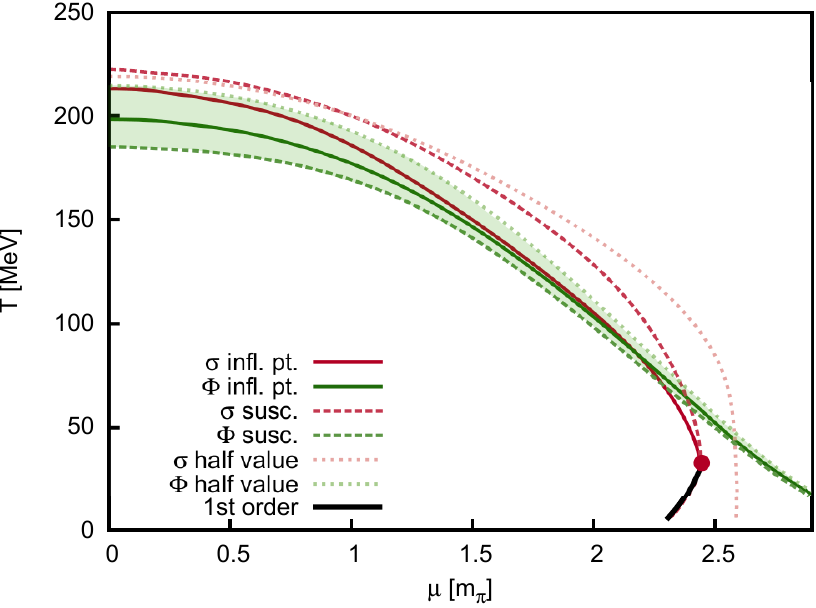}
    \caption{PQMD model phase diagram for an $O(6)$ symmetric effective potential and $T_0=T_0(\mu)$ ($b=2.60$). Color coding as in Fig.~\ref{fig:pqmdd0nomu}.}
    \label{fig:pqmdd0withmu}
   \end{figure}

Two phase diagrams obtained from calculations with $O(6)$-symmetric
effective potential are shown for comparison  
in Figs.\ \ref{fig:pqmdd0nomu} and \ref{fig:pqmdd0withmu}. In
Fig.~\ref{fig:pqmdd0nomu} we have used a constant $T_0 = 212$ MeV in
the Polyakov loop potential, while Fig.~\ref{fig:pqmdd0withmu} shows
the corresponding result with the chemical-potential-dependent
$T_0(\mu) = T_0 \exp\{- c\, \mu^2/T_0^2\} $ and $c=0.46 $ corresponding
to  $b=2.6$ in Eq.~(\ref{eq:T0scaling}) to model the leading finite
density effects from Debye screening via the assumption of constant
effective charge along the transition line in the pure glue potential
as discussed in Subsection \ref{subsec:pqmd}.
In the first case the 
deconfinement temperature only shows a very slight decrease with
increasing chemical potential. The three different definitions are
used to visualize the width of the deconfinement crossover. In the
second case one observes a considerable decrease of the deconfinement
crossover temperature with chemical potential. At the same time, as one can
infer directly from Fig.~\ref{fig:condensatesmu0} and indirectly from the
focusing of the deconfinement crossover lines corresponding to
different crossover criteria, the crossover becomes 
increasingly rapid with increasing chemical potential but remains a
continuous transition throughout. The corresponding chiral crossover
lines remain more or less parallel to the deconfinement transition 
up to a temperature of around 80~MeV from where on the chiral transition
bends downwards and eventually merges into the critical endpoint, whereas the
deconfinement crossover continues to decrease approximately 
linearly with the chemical potential until it starts bending away from
the $T=0$ axis. Apart from this splitting of the two transitions near
the critical endpoint, which was not observed in the analogous
3-color calculations, these phase diagrams agree qualitatively with
the corresponding PQM model results for QCD
\cite{Herbst:2012ht,Herbst:2013ail}. 
In particular, the density-dependent transition temperature $T_0(\mu)$
in the Polyakov loop potential has the same overall effect in either
case. Moreover, the phase diagrams in Figs.~\ref{fig:pqmdd0nomu} and
\ref{fig:pqmdd0withmu} both show quarkyonic 
phases of confined but chirally restored matter although their sizes
differ considerably. One should keep in mind, however, that both
phase diagrams yield equally inappropriate descriptions of two-color QCD at finite
baryon density as we have so far neglected the diquarks as the baryonic
degrees of freedom in this theory.

\subsection{Full effective potential}
\label{subsec:fullflow}
\begin{figure}[!ht]
	\subfigure[$T_0=\text{const.}$ ($b=0$)]{
    \includegraphics[width=0.97\columnwidth]{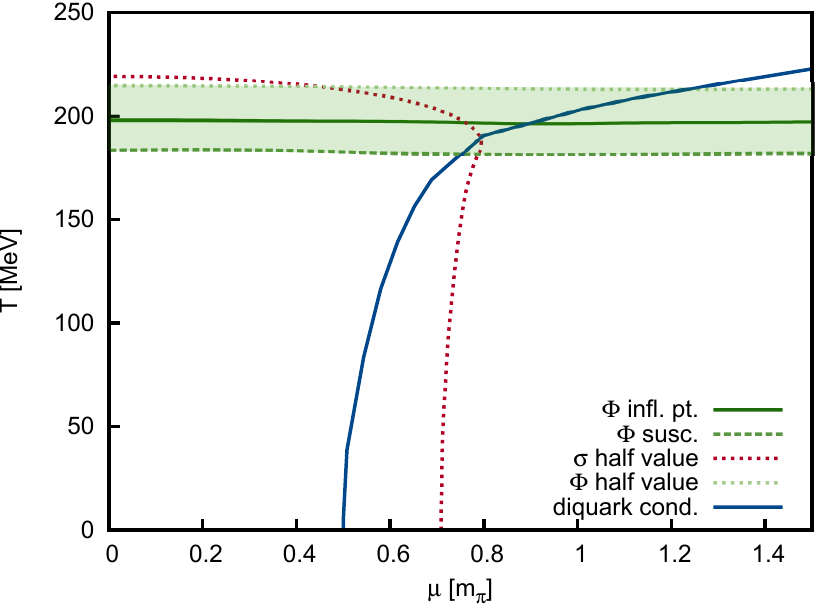}
	\label{fig:pqmdnomu}
	}
	\subfigure[$T_0=T_0(\mu)$ ($b=2.60$)]{
    \includegraphics[width=0.97\columnwidth]{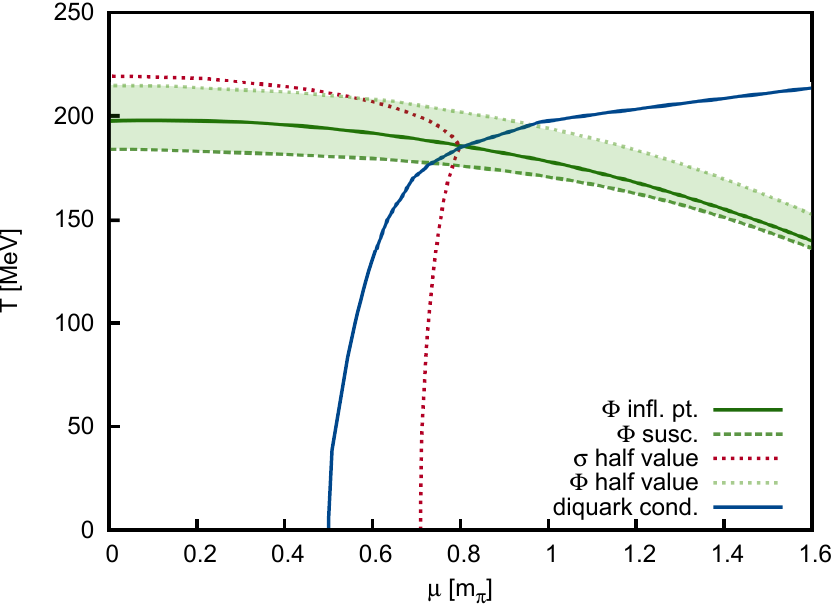}
	\label{fig:pqmdwithmu}
	}
	\subfigure[$T_0=T_0(\mu)$ ($b=5.20$)]{
    \includegraphics[width=0.97\columnwidth]{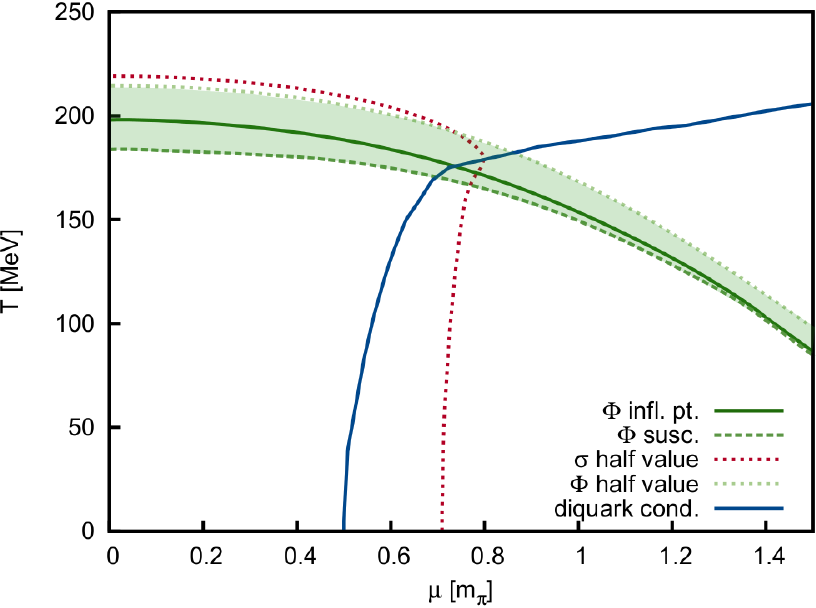}
	\label{fig:pqmdwithmubeta2}
	}
\caption{PQMD model phase diagrams illustrating the effects of matter-backcoupling using different parameter values for $b$, see Eq.~(\ref{eq:T0scaling}). Chiral crossover lines (half value of the condensate) are depicted in red, deconfinement crossover lines in green (comparing three different crossover criteria) and the second order phase boundary of the diquark condensation phase in blue.}
   \end{figure}  

The correct inclusion of the diquark degrees of freedom is addressed
in the present subsection where we discuss the full solution of the
PQMD model flow equation for the effective potential. Again we compare
the phase structure for a constant $T_0$ in
Fig.~\ref{fig:pqmdnomu} to that with the chemical-potential-dependent
$T_0=T_0(\mu)$ in Fig.~\ref{fig:pqmdwithmu} and Fig.~\ref{fig:pqmdwithmubeta2}. The parameters are the
same as in the previous subsection. Similar to our observation there,
without the density dependence from the Debye mass in the pure glue
potential, the deconfinement crossover is almost independent of the chemical
potential here as well. This is fully in line with previous PNJL model
mean-field results for constant $T_0$ \cite{Brauner:2009gu}. 

The lines from inflection points and susceptibility peaks for the
chiral transition (not shown here) both stay above the   
diquark condensation phase boundary and lose their meaning as
pseudo-critical lines at large chemical potentials. 
That is why we only show the half-value line as a representative
contour line to indicate chiral symmetry restoration, in particular
inside the diquark condensation phase where it is related to the
analogue of the BEC-BCS crossover in two-color QCD. 

Very similar to the results from the previous section the inclusion of
matter backcoupling on the gauge sector, such as the density-dependent 
Debye screening, lead to a decrease of the deconfinement
crossover temperature with increasing chemical potential in
Fig.~\ref{fig:pqmdwithmu} and Fig.~\ref{fig:pqmdwithmubeta2}. 
As a result, the deconfinement transition
traverses deep into the diquark condensation phase leading to a phase
diagram which is in overall good qualitative agreement with recent
lattice results \cite{Boz:2013rca,Cotter:2012mb}. The comparison between 
Fig.~\ref{fig:pqmdwithmu} and Fig.~\ref{fig:pqmdwithmubeta2} serves to 
illustrate the impact of the non-perturbative coefficient $b$. As expected,
increasing $b$ leads to a stronger decrease of the deconfinement
crossover temperature with increasing chemical potential. This comes
together with a certain suppression of the diquark condensation
transition temperature above their intersection point. The shape of
the diquark condensation phase tends to become more rectangular, which
would be in quite good agreement with most recent lattice results
\cite{Boz:2013rca}.

On a more quantitative level, however, these lattice results indicate that the 
practically $\mu$-independent horizontal boundary of the diquark
condensation phase occurs  at a temperature which is only
about half of that of the deconfinement transition at $\mu = 0$
\cite{Boz:2013rca}. This might at least partially be explained by
 the rather heavy quark masses there, which should be further
 investigated, but at the moment it is nevertheless 
at odds with the available model results. 

Disregarding the differences in physical parameters between the
lattice simulations and the FRG calculation, the deconfinement
crossover in the lattice simulations still shows a somewhat sharper decrease
with chemical potential as compared to the FRG calculation. This might
hint at effects of matter backcoupling on the gauge
sector at large chemical potentials which are beyond 
the expansion in (\ref{eq:bmuexpansion}) for the leading
order $\mu^2$-dependence in $T_0(\mu)$ from the density-dependence of
the Debye mass. Furthermore, previous
4-flavor lattice simulations \cite{Kogut:2001if} have provided
evidence of a first order finite temperature phase transition at large
chemical potentials implying the existence of a tricritical point
along the diquark condensation phase boundary which was attributed to light
mesonic/diquark fluctuations \cite{Kogut:2001if}. While this will certainly depend
on the number of flavors, the fact that it is not observed in
our 2-flavor FRG calculations here, although the latter includes the relevant fluctuations,
points to insufficiencies at large chemical potentials which might be
resolved by considering chemical-potential-dependent initial conditions.
More generally, from a QCD perspective one should take into
account the temperature and chemical potential dependence 
of the two-gluon exchange diagrams which drive the flow of 4-Fermi
interactions at large scales. The difference of the vacuum contribution and the
corresponding contributions at finite temperature and chemical
potential translate after bosonization into temperature- and
chemical-potential-dependent contributions to the UV potential. 

At some point, however, a proper inclusion of gauge degrees of freedom
beyond the simple coupling to a phenomenological Polyakov
loop potential becomes indispensable. In a functional approach such a
description of two-color QCD could be achieved quite analogously to what
has been done in the case of three colors \cite{Braun:2009gm} already.   

\section{Summary and Conclusions}
In this paper we presented first results on the phase diagram of
two-color QCD from a QMD model calculation on the one hand extending
earlier results \cite{Strodthoff:2011tz} by the inclusion of gauge
degrees of freedom in the form a coupling to phenomenological Polyakov
loop potential, and on the other hand extending earlier mean-field
calculation \cite{Brauner:2009gu} by the inclusion of collective
mesonic and baryonic excitations and effects of matter-backcoupling on
the gauge sector. Furthermore the results presented here include
additional UV contributions ensuring thermodynamic consistency which
are particularly important in regions of the phase diagram where the
temperature is large compared to the UV cutoff scale. These lead to a
certain decrease of both, the chiral and the diquark-condensation
transition temperatures.

Similar to the 3-color case \cite{Schaefer:2007pw,Herbst:2010rf}, the
matter backcoupling onto the gauge sector, here implemented via a
chemical-potential-dependent temperature $T_0(\mu)$ entering the
Polyakov loop potential, turns out to be crucial for the 2-color case
as well. Whereas the deconfinement transition temperature stays
practically independent of the chemical potential for constant $T_0$,
similar to what has been observed in a PNJL model analysis
\cite{Brauner:2009gu}, the matter backcoupling leads to a significant
decrease of the transition temperature with increasing chemical
potential. The corresponding phase diagram is found to be in good
qualitative agreement with recent lattice results.

It would provide interesting insights to study the PQMD model in an
extended truncation which goes beyond the zeroth order in the
derivative expansion and which is fully consistent with the reduced
symmetry of the theory at vanishing chemical potential, although this
is expected to lead only to quantitative changes of the phase
diagram. The resolution of the remaining discrepancies discussed in
the previous section will be important for our general understanding
of the reliability of FRG results at large chemical potentials and
might in this sense also be directly relevant for the corresponding
3-color calculations. Further input might come from the comparison of
two-color QCD thermodynamics between lattice results and functional
methods. In this respect we look forward to further lattice results
for two-color QCD with $N_f=2$ quark flavors which can be most easily
treated in the effective model and functional renormalization group 
approaches as discussed here.

\section*{Acknowledgements}
The authors thank Jan Pawlowski for insightful discussions. This work
was supported by the Helmholtz International Center for FAIR within
the LOEWE initiative of the State of Hesse, and by the European
Commission, FP7-PEOPLE-2009-RG, No. 249203. N.S. is supported by the grant ERC-AdG-290623.
 \bibliographystyle{h-elsevier.bst} 
 \bibliography{pqmd}
      	 
\end{document}